%% file: main.tex
\documentclass[sigconf]{acmart}
\AtBeginDocument{%
  }

\acmConference[TOCHI]{}{2025}{ACM}

\setcopyright{cc}
\setcctype{by-sa}
\acmJournal{TOCHI}
\acmYear{2025} \acmVolume{1} \acmNumber{1} \acmArticle{1} \acmMonth{1} \acmPrice{}\acmDOI{10.1145/3731756}

\usepackage{enumitem}
\usepackage{xspace}
\usepackage[capitalise, noabbrev]{cleveref}
\usepackage{multirow}
\usepackage{wrapfig}

\newcommand{\sysname}{\textsc{ROPE}\xspace}
\newcommand{\sysgroup}{\texttt{\sysname}\xspace}
\newcommand{\ctrlgroup}{\texttt{PE}\xspace}
\newcommand{\paragraphBold}[1]{\paragraph{\emph{\textbf{#1}}}\xspace}
\newcommand{\task}[1]{\texttt{#1}\xspace}
\newcommand{\req}[1]{\texttt{``#1''}\xspace}
\definecolor{cquote}{HTML}{3c4043}
\newcommand{\quoteinline}[1]{{\color{cquote}{``#1''}\xspace}}

\begin{document}

\title[\sysname: Requirement-Oriented Prompt Engineering]{What Should We Engineer in Prompts? Training Humans in Requirement-Driven LLM Use}

\author{Qianou Ma}
\email{qianouma@cmu.edu}
\affiliation{%
  \institution{Carnegie Mellon University}
  \city{Pittsburgh}
  \state{PA}
  \country{USA}
}

\author{Weirui Peng}
\email{wp2297@columbia.edu}
\affiliation{%
  \institution{Columbia University}
  \city{New York}
  \state{NY}
  \country{USA}
}

\author{Chenyang Yang}
\email{cyang3@cs.cmu.edu}
\affiliation{%
  \institution{Carnegie Mellon University}
  \city{Pittsburgh}
  \state{PA}
  \country{USA}
}

\author{Hua Shen}
\email{huashen@uw.edu}
\affiliation{%
  \institution{University of Washington}
  \city{Seattle}
  \state{WA}
  \country{USA}
}

\author{Kenneth Koedinger}
\email{koedinger@cmu.edu}
\affiliation{%
  \institution{Carnegie Mellon University}
  \city{Pittsburgh}
  \state{PA}
  \country{USA}
}

\author{Tongshuang Wu}
\email{sherryw@cs.cmu.edu}
\affiliation{%
  \institution{Carnegie Mellon University}
  \city{Pittsburgh}
  \state{PA}
  \country{USA}
}

\renewcommand{\shortauthors}{Ma, et al.}

\begin{abstract}
Prompting LLMs for complex tasks (e.g., building a trip advisor chatbot) needs humans to clearly articulate customized requirements (e.g., ``start the response with a tl;dr''). 
However, existing prompt engineering instructions often lack focused training on requirement articulation and instead tend to emphasize increasingly automatable strategies (e.g., tricks like adding role-plays and ``think step-by-step'').
To address the gap, we introduce Requirement-Oriented Prompt Engineering (\sysname), a paradigm that focuses human attention on generating clear, complete requirements during prompting. We implement \sysname through an assessment and training suite that provides deliberate practice with LLM-generated feedback. 
In a randomized controlled experiment with 30 novices, \sysname significantly outperforms conventional prompt engineering training (20\% vs. 1\% gains), a gap that automatic prompt optimization cannot close. Furthermore, we demonstrate a direct correlation between the quality of input requirements and LLM outputs. Our work paves the way to empower more end-users to build complex LLM applications.
\end{abstract}



\begin{teaserfigure}
    \centering
\includegraphics[trim={0cm 5cm 6cm 0cm}, clip,width=\linewidth]{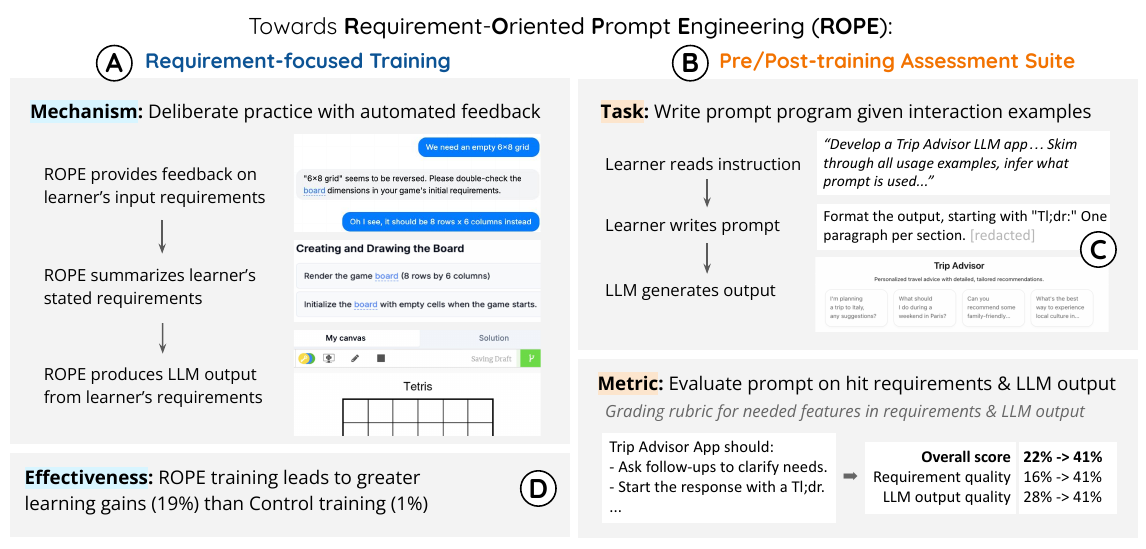}
    \vspace{-20pt} 
  \caption{We propose Requirement-Oriented Prompt Engineering (ROPE) and contribute a training and assessment suite for ROPE. (A) ROPE training helps learners write effective prompt programs by providing automated feedback on adding and clarifying requirements. (B) In our pre-post assessments, we ask learners to write prompts to develop customized applications (e.g., Trip Advisor via a prompt in C) and assess learners’ prompt quality on both requirement quality and LLM output quality. (D) In a pre-post randomized experiment, ROPE training significantly improves learners’ prompt quality, compared to conventional prompt engineering training. }
  \Description{Diagram showing the ROPE (Requirement-Oriented Prompt Engineering) framework. It includes requirement-focused training with feedback mechanisms and pre/post-training assessment tasks. The training provides feedback, summarizes requirements, and evaluates prompt output. A study result shows increased learning gains with ROPE training compared to control.}
  \label{fig:teaser}
\end{teaserfigure}

\received{08 Dec 2024}
\received[revised]{04 Apr 2025}
\received[accepted]{15 Apr 2025}

\maketitle

\input{sections/intro}
\input{sections/related}
\input{sections/rope_paradigm}
\input{sections/training_design}

\input{sections/user_study}
\input{sections/user_study_result}
\input{sections/discussion}

\section{Conclusion}
In this work, we advocate for focusing prompt engineering effort on human-centered tasks, ensuring users include all necessary requirements in their prompt to achieve goals. We introduce the Requirement-Oriented Prompt Engineering (\sysname) paradigm, and design training materials where prompting novices practice requirement articulation on complex prompting tasks. We also develop aligned assessment metrics capturing both the intrinsic quality of user prompts in terms of requirement quality, as well as the extrinsic quality of prompt effectiveness in achieving intended outcomes.
By providing targeted feedback on requirements, our \sysname system helps users produce higher-quality requirements and prompts more effectively than traditional prompt engineering training.
As we look to the future, it is clear that the demand for LLM-based applications will only grow. 
As LLMs become more integrated into complex task-solving for more users, the ability to clearly articulate requirements will be key to effectively guiding LLMs.
We believe that users should be equipped with the foundational skills to prepare for the \sysname future, and the right training will empower users to harness the full potential of LLMs.

\begin{acks}
Thanks to all the participants for the pilot, interview, and user study in this work. Thanks to Kelly Rivers, Michael Taylor, Michael Hilton, Michael Xieyang Liu, Xinran Zhao, Lauren Sands, Austin Schick, David Kosbie, and Daniel Anderson for all the insights, advice, and help. Thanks to the CMU HCII faculty and Ken's lab for feedback. Thanks to the National Science Foundation (award CNS-2213791, 2414915) and gift fund from Google for support of this work. Thanks to the OpenAI research credit program and Amazon AI research gift fund. Thanks to all reviewers of this work.
\end{acks}

\bibliographystyle{ACM-Reference-Format}
\bibliography{paperpile, ref}


\end{document}

%% file: sections/intro.tex
\section{Introduction}
\label{sec:intro}

General-purpose AIs like large language models (LLMs) have evolved from simple next-word predictors~\cite{brown2020language} to powerful assistants capable of fulfilling complex user needs~\cite{Qin2024-mn, ouyang2022training}.
This improvement in LLM's ability to follow instructions has encouraged users to delegate increasingly intricate tasks to these models. 
In the early days of prompt engineering, users primarily focused on refining the wording of simple instructions to improve LLM output quality~\cite{jiang2020can, shin2020autoprompt}.
Today, prompts resemble detailed ``essays'' that define LLM behaviors.
Rather than one-off small requests, these prompts start to power the generation of complex {applications}, and \emph{everyday users} can write complex prompts to either generate code for softwares~\cite{Wodecki2024-hz, yang2024sweagent} or tailor general-purpose LLMs into special purpose LLM applications~\cite{zhang2024first}.
For instance, an LLM application (or a GPTs) like \task{Trip Advisor}\footnote{\url{https://chatgpt.com/g/g-K38J9feSL-trip-advisor}} can be built solely using prompts similar to \cref{fig:teaser}C. 
These LLM applications function similarly to traditional software and are accessible through LLM App Stores like GPT Store\footnote{\url{https://openai.com/index/introducing-the-gpt-store/}} and FlowGPT\footnote{\url{https://flowgpt.com/}}, where they can be reused with a single click. In just a few months, over 30,000 ``LLM app developers'' have created millions of GPT apps, with the most popular ones being used more than 5 million times~\cite{zhang2024first}. This trend signals a future of \textbf{end-user programming through natural language}, where anyone can create \textit{reusable programs} just by writing prompts \cite{Feng2024-rj}.

\begin{figure}[t!]
\includegraphics[width=1\linewidth, trim={0 3.2cm 13.5cm 0cm},clip]{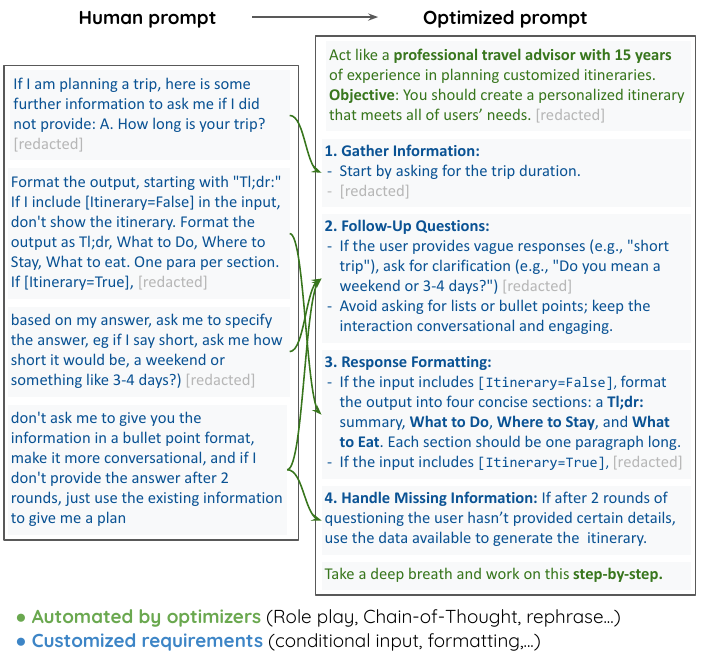}
\caption{
Automatically refine a user's prompt for \task{Trip Advisor} using the optimizer Prompt Maker. Customized requirements still need to come from the human prompt, while the optimizer improves on non-requirement prompting aspects such as  fluency, role plays, structures, etc.}
\label{fig:motivation}
\Description{Side-by-side comparison of a human-written prompt and its optimized version for a Trip Advisor task. The human prompt is informal and fragmented, whereas the optimized prompt is structured with sections. The optimized version improves fluency and structure using role-play and chain-of-thought methods.}
\end{figure}

The future of writing prompt programs looks promising, but \emph{status quo} prompt engineering remains a challenge. 
Various studies show that optimizing LLM outputs requires well-rounded prompts, involving fluent writing, clear instructions, personas, formats, and effective adaption of prompting techniques like Chain-of-Thought~\cite{bsharat2023principled, Mesko2023-um, Marvin2024-iv, Schmidt2024-bh}. 
These complexities often confuse novices, causing them to make \emph{ad hoc} revisions to prompts without a clear understanding of what needs improvement~\cite{Zamfirescu-Pereira2023-is}.
Recognizing the challenge, the NLP community is developing \emph{prompt optimizers} to automate some prompt refinement steps, e.g., incorporating role-plays, applying effective techniques, structuring prompts, and enhancing text fluency (e.g., \cref{fig:motivation}, discussions in \cref{subsec:PO}).

The rise of optimizers makes us think: What aspects of prompt engineering should humans still master?  
We argue that the remaining part of manual prompt creation is akin to the core step before engineering in a design process: \textbf{requirement specification}. A \emph{requirement} is a description of user goals; it defines and documents how a system should behave, including expected inputs and outputs~\cite{lamsweerde2009requirements} (e.g., \req{Start the response with a tl;dr} for \task{Trip Advisor} in \cref{fig:teaser}B).
Requirements are the backbone of program functionalities, and user-specified requirements in prompts (e.g., \cref{fig:motivation}) are the driving forces behind complex LLM adaptations.
\textbf{Customized requirements are hard to predict, and thus difficult to fully automate and must be specified by humans.}
We argue that in a future where prompt programs are common, individuals should develop proficiency in requirement articulation. In other words, \textbf{we need to shift our focus towards Requirement-Oriented Prompt Engineering, an emerging paradigm that we call ``\sysname''}.

Articulating clear and complete requirements is a known challenge in fields such as product design and software engineering~\cite{Liu2023-hf, Nam2024-td, Wang2009-sc}, and poor requirements often cause inefficiency, frustrations, and software failures~\cite{Zamfirescu-Pereira2023-is, millett2007software}. This difficulty is amplified by the unpredictable nature of LLMs, which complicates controlling outputs~\cite{Arzberger2024-xu, Chopra2023-tb}. Even experts need multiple iterations to refine requirements~\cite{Shankar2024-hy}, and novices frequently make requirement mistakes~\cite{babe2023studenteval}. Recognizing both the importance and difficulty of this task, we pose the research question: \textbf{Can we help end-users better instruct LLMs to achieve their goals through \sysname training?}

To answer this question, we present the \sysname paradigm as a future human-LLM partnering strategy (\cref{sec:paradigm}), and we develop requirement-focused training and assessments to (1) help users practice writing complete and correct requirements, and (2) deepen our understanding of how requirements affect the quality of LLM outputs.
Specifically, we formalize \emph{tasks} (\cref{subsec:design-task}) with complex requirements (e.g., game development), and design corresponding \emph{assessments} (\cref{subsec:design-assessment}) that can measure the requirement quality, the LLM output quality, and their correlations.
We also design and implement an interactive training mechanism (\cref{subsec:design-interaction}) that 
\emph{disentangles requirement articulation from other adjacent prompting aspects}. Our system supports deliberate practice on requirement refinement by generating immediate feedback grounded on reference requirements.

We conduct a randomized controlled user study (\cref{sec:user-study-design}) with 30 prompting novices, comparing the prompts users write before and after training, and against a control group that received standard prompt engineering training.
We confirm that \emph{our requirement-focused training is effective} (\cref{sec:user-study-result}):  
novices receiving this training significantly improved their requirement performances by {two-fold} in the pre-to-post test, achieved {significantly more gains} than conventional prompt engineering training, and learned to iterate on prompts in a more structured way towards requirements. 
Moreover, we confirm that \emph{requirements are indeed important for the LLM generation}: There is a {strong} positive correlation between the requirement qualities and the LLM output qualities.
Importantly, \emph{the training gains are complementary to potential improvements from optimizers and models}: 
Using a popular prompt optimizer, we observe that optimizers can indeed improve requirement quality, but they cannot close the gap between the baseline and the training groups; Moreover, when switching from \texttt{GPT-4o} to the more advanced reasoning model \texttt{o3-mini}, the model's improved instruction-following capabilities made its outputs more reflective of users' explicit requirements, rather than rendering requirements obsolete. 
We conclude with an in-depth discussion (\cref{sec:discuss}) on future directions for human training and optimizer design to enhance human-LLM complementarity in \sysname. 

In summary, we make the following contributions:

\begin{itemize}[labelwidth=*,leftmargin=1.3em,align=left, topsep=0em]
\item We identify the significance of requirement-based prompting, and propose the \sysname paradigm as a foundation for future human-LLM task delegation in prompt engineering. 
\item To the best of our knowledge, we are the first to investigate training specifically for requirement-based prompting. We developed tasks, assessments, and training systems that effectively improve requirement generation skills. 
\item We provide both quantitative and qualitative insights for the future of prompting, highlight the relationship between the quality of requirements and LLM performance across model advancements, and discuss current optimizers' limitations and potentials. 
\end{itemize}

As natural language prompt programs become more widespread, everyone needs the skill to write good requirements. We believe that with our \sysname paradigm, LLMs can empower anyone to program reusable AI assistants.
We open-source our training system, supplemental study materials, and user prompts at \url{https://github.com/mqo00/rope/}.

%% file: sections/related.tex
\section{Related Works}
\label{sec:relate}

We review three key areas relevant to our work: prompt engineering, prompt optimization in natural language processing, and requirements engineering in end-user software engineering. 
We highlight a significant gap in the literature: the lack of requirement-oriented prompt engineering training for end users, which we propose to address in this paper.

\subsection{Current Prompt Engineering Practices and Challenges}
\label{subsec:PE}

Prompt engineering (PE) has been essential for crafting effective input instructions to guide LLMs toward generating desired outputs~\cite{Liu2023-oo, Bach2022-xu, Zamfirescu-Pereira2023-is}. 
However, the non-deterministic nature of LLMs makes it challenging for humans to predict LLMs' behavior across prompts~\cite{ouyang2023llm, Denny2024-ft, Sun2022-af}. This leads to wasted time on unproductive strategies, such as trivial wording changes \cite{Feldman2024-jc, Denny2024-ft, Zamfirescu-Pereira2023-is}. 
While some challenges can be mitigated through automation (see \cref{subsec:PO}), e.g., automating word choices, human requirements (some also refer to as constitutions \cite{petridis2024constitutionmaker} or principles \cite{louie2024roleplay} in literature) remain crucial for customized or specialized tasks \cite{Liu2023-oo}. 
We define these tasks as \emph{LLM-hard}, as a simple prompt cannot produce a satisfactory response. 
Developing reusable prompts for customized chatbots or GPTs \cite{Bach2022-xu, Zamfirescu-Pereira2023-is} is a common {LLM-hard} task. While a \textit{prompt} is an input to LLMs, a \textit{reusable prompt} is a set of instructions for recurring goals, essentially functioning as a \textit{prompt program}. 
As more users create reusable prompts, the ability to write good prompt programs with clear requirements becomes increasingly important.

However, prompt creation can be complex and inaccessible to non-experts who lack technical knowledge of LLMs, highlighting the need for more end-user prompt engineering (EUPE) scaffolding and training \cite{Zamfirescu-Pereira2023-is}. 
Various strategies have been developed to support PE, such as prompt technique catalog~\cite{bsharat2023principled, Liu2022-ac, Wang2024-xj}, chat-based tools like GPT-builder,\footnote{\url{https://chatgpt.com/create}} and toolkit for orchestrating complex tasks~\cite{Chase_LangChain_2022, khattab2023dspy}. 
Each tool has trade-offs, balancing flexibility, precision, and technicality. For example, chat-based tools may ask narrow clarifications without fully understanding the context, leaving users to self-refine requirements when LLM responses are open-ended \cite{Lahiri2022-gb, Qian2024-ok}.

Existing PE training emphasizes the importance of explicitly stating requirements within prompts \cite{Mesko2023-um, Schmidt2024-bh, Santana2024-bg}, yet there is still a lack of focused training on requirement articulation skills for everyday users. 
Without training, non-experts often make mistakes such as missing requirements and writing conflicting requirements in prompts \cite{Feldman2024-jc, Nam2024-td, Prather2024-wc, Mahdavi-Goloujeh2024-wh}. Even experts need many iterations to improve their requirements for complicated LLM tasks \cite{Shankar2024-hy}. However, training novices on PE is difficult; for instance, novice programmers may not improve at prompting within a 75 minute study when provided with test cases and generated code as feedback \cite{Nguyen2024-sd}.
We aim to address this gap by proposing a \sysname training for end-users to improve their requirements engineering skills in the context of prompt creation.

\subsection{Instruction Following for Foundation Models}
\label{subsec:PO}

Optimizing LLM performance on customized tasks typically follows two approaches: improving the models directly, or refining prompts with models fixed. 
The former has enabled the versatility of models like GPT-4,\footnote{\url{https://openai.com/index/gpt-4/}} Gemini,\footnote{\url{https://gemini.google.com/}} and LLaMA 3,\footnote{\url{https://www.llama.com/}} which are fine-tuned through instruction tuning~\cite{ouyang2022training} and further aligned with human preferences using Reinforcement Learning from Human Feedback (RLHF)~\cite{christiano2017deep}.
For the latter, the NLP community has been exploring LLMs' capabilities to follow requirements (or ``constraints'') with various techniques such as decomposition (least-to-most prompting \cite{Zhou2022-vv} and AI chains \cite{Wu2022-bn}) for complex tasks. Various datasets have been proposed to assess whether LLMs can fulfill constraints, such as IFEval~\cite{Zhou2023-tb}, INFOBench~\cite{Qin2024-mn}, and FollowBench~\cite{Jiang2024-ia}. These datasets focus on evaluating models' abilities to handle compositional constraints, and have demonstrated promising advancements.

Our work hypothesizes that requirements can be central to human-LLM interactions, as we have observed the LLM proficiency in following instructions.
However, existing requirement datasets are often synthetic, containing prompts that are intentionally tricky or unrealistic, with requirements generated automatically via templates and limited to certain categories (e.g., sentence or paragraph length constraints, format constraints like output in JSON). While useful for identifying \emph{LLM-hard} problems for requirement-focused training, these datasets do not fully capture how LLMs respond to real-world user prompts. Human prompts may have more diverse requirements across various tasks and less standardized language --- a gap we seek to fill in this paper.

Building on this foundation of instruction following, \emph{prompt optimizers} have been proposed to improve the performance of frozen LLMs on user-defined downstream tasks by automatically refining user-written prompts. 
Automation ranges from simple adjustments like adding role-playing or chain-of-thought prompting~\cite{wei2022chain} (e.g., Prompt Maker in \cref{fig:motivation}) to more advanced methods like TextGrad~\cite{Yuksekgonul2024-wn}, which searches for optimal wording, PromptCharm~\cite{Wang2024-xj}, which enables users to adjust diffusion models' attention to keywords in prompts, and DSPy~\cite{khattab2023dspy}, which extends the search space to few-shot prompts and various prompting techniques.
While these sophisticated methods make more tailored changes, they are harder to apply and often require access to labeled datasets. 
Despite being in early stages, these optimizers aim to relieve humans from exhausting all possibilities in the natural language prompting space, as automated experiments with semantically-preserving edits can be more effective. We share their vision and foresee a future where optimizers play a crucial role in human-LLM interaction.

\subsection{Requirements in Software Engineering and Design}
\label{subsec:RE}

In software engineering and end-user software engineering, \textit{requirement} describes what a human wants to achieve and how a program should behave in the world \cite{Ko2011-kn}.
In product and engineering design, \textit{requirement} is a description of the desired solution to a problem \cite{McDonnell2014-tl, Wang2009-sc}.
Requirements encompass all the necessary conditions, constraints, and desired outcomes to ensure the output aligns with the human's needs and expectations, and specifying requirements is often a crucial step in both the design and engineering process \cite{Dubberly2004-ce}.
\emph{Requirements engineering} stems as a field that focuses on generating and documenting requirements for software \cite{Walia2009-pk}. Good quality requirements need to be \textit{accurate} and \textit{complete}, without \textit{commission} (inclusion of irrelevant or incorrect details) and \textit{omission} (exclusion of necessary details) defects \cite{Alshazly2014-qr}. 
Previous studies have identified requirements engineering as a challenging skill to master. While training mechanisms have been developed to help students avoid commission and omission errors \cite{Neto2013-ij}, a significant skill gap remains between graduates and professional engineers \cite{Radermacher2014-be, Eckerdal2006-qz}.

Parallels can be drawn between prompt engineering (PE) and requirements engineering. For example, adapting LLMs to diverse scenarios demands well-specified requirements, and prompt authors often need to iterate on prompts that are too ambiguous for the LLM to interpret~\cite{petridis2024constitutionmaker}. 
While PE training stresses the importance of clear requirements~\cite{Mesko2023-um, Schmidt2024-bh, Santana2024-bg}, this is often presented just as one of many prompting principles (e.g., one of 26 \cite{bsharat2023principled}), limiting its impact on prompt authors.
In fact, little attention has been given to explicitly training end-users on requirements engineering in the context of prompt programs, including both natural-language-to-code and natural-language driven GPT applications. 
While several tools like EvalLM~\cite{Kim2024-lu} and EvalGen~\cite{Shankar2024-hb} have explored extracting user criteria to support prompt \textit{evaluation}, the concept of requirements has not yet been fully emphasized in the context of prompt \textit{construction}.
We aim to address this gap by introducing the \sysname paradigm and developing a requirement-focused training mechanism to help end-users more effectively instruct LLMs to achieve their desired goals. 

%% file: sections/rope_paradigm.tex
\section{The \sysname Paradigm}
\label{sec:paradigm}

We offer our definition of Requirement-Oriented Prompt Engineering (\sysname) and describe why, when, and who need it, and why we propose a training toward \sysname in this work. 
\sysname is a human-LLM partnering strategy where humans maintain agency and control of the goals via specifying requirements for LLM prompts.
\sysname represents an \textbf{emerging paradigm in how we interact with LLMs, focusing on the importance of crafting accurate and complete requirements to achieve better results, especially for complex, customized tasks.}

\begin{figure}
\includegraphics[width=1\linewidth, trim={1 10.5cm 11.5cm 0cm},clip]{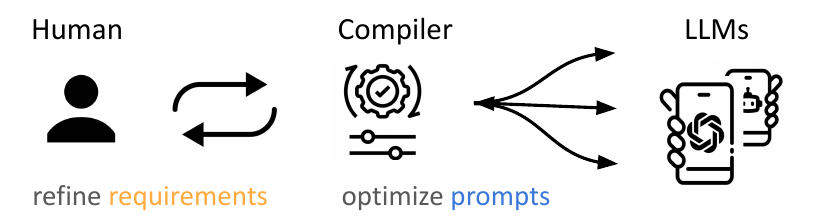}
\caption{Our envisioned \sysname paradigm.}
\vspace{-15pt}
\label{fig:paradigm}
\Description{A conceptual diagram of the ROPE paradigm. A human refines requirements, which are passed to a compiler that optimizes prompts for large language models (LLMs). Arrows indicate the flow between human and compiler and from compiler to LLMs.}
\end{figure}

\paragraphBold{The definition of requirements}
We define a \emph{requirement} as a skeleton instruction that communicates an essential condition or constraint on desired LLMs output (e.g., \req{Response is less than 100 words}). Requirements instruct LLMs to perform tasks that may deviate from their default behavior, guiding LLMs to align with user's goals.

In our work, we focus on evaluating requirements quality rather than quantity, and we operationalize requirement quality in our work (defined in \cref{subsec:design-assessment}) by adopting a taxonomy from the requirements engineering literature (\cref{subsec:RE}). 
Ensuring requirements are accurate and complete is essential for effective LLM prompting, as poor-quality requirements can lead to harmful outcomes --- e.g., missing requirement \req{anonymize the data} may make LLMs keep identifiable information in data, and vague requirement \req{delete harmful content} without a clear definition of ``harmful'' may cause LLMs to misinterpret sensitive topics like mental health or race as harmful.

Here we also focus on \emph{natural language requirements} due to its universal understanding and tight connection to LLM prompting. While multi-modal requirements can be compiled into prompts for different generative AI models, we leave this exploration for future work.

\paragraphBold{The relation between requirements and prompts}

We view a prompt as a \emph{super-set} of requirements. It contains not only users' customized requirements, but also other (orthogonal) factors like text fluency and standard prompting tricks.
Among these factors, requirements are more \emph{user-centered} and \emph{LLM-agnostic} --- users' goals generally remain consistent across models (e.g., GPT-4, Gemini).
Fully articulating these requirements is essential, as omitted customized details can be difficult to recover automatically (explored further below). 
In contrast, other factors tend to be more \emph{LLM-centered}: different models may prefer different writing styles, example types, or requirement order. 
While optimizing these factors can enhance LLM performance, they are better suited for automation rather than humans due to their (1) formulaic nature (e.g., wrapping prompts in a template consistently improves Llama-3-8b's performance on any task~\cite{dubey2024llama}) and (2) idiosyncratic behavior (e.g., GPT-3 can be sensitive to semantically invariant edits that humans do not expect to make drastic differences~\cite{Zamfirescu-Pereira2023-is}). 
This division is crucial given the rapid evolution of LLMs, as developing a mental model for a specific LLM can be a waste of effort in the long run \cite{Liu2022-ac}.

With this distinction, we view \sysname as a task-delegation strategy. In crafting executable prompt programs, users should focus on \emph{iterating and refining requirement-related components}, leaving other aspects automated. We find recent advancements in prompt optimizers (\cref{subsec:PO}) particularly promising. 
With a consistent set of user requirements, we hope future optimizers could function like program synthesizers, searching through possible ``implementations'' --- in terms of phrasing, examples, and structure --- to generate prompts that best align with each LLM’s preferences.

\paragraphBold{The applicability of \sysname: LLM-hard tasks and prompt programmers}

In theory, any task can have countless requirements; for instance, implicit needs like \req{reply in the same language} are almost always assumed in LLM applications. However, not all requirements need to be explicitly stated. 
For standard tasks well-represented in LLM training data, e.g., \req{correct the grammar in this email}, LLMs can meet expectations relying on their parametric knowledge without explicit clarification (e.g., on what grammatical rules exist).

We argue that explicitly articulating requirements becomes more crucial for what we refer to as \textbf{LLM-hard} tasks.
These tasks require significant customization where users must intentionally \emph{deviate LLMs from their default behaviors}~\cite{Qin2024-mn, Shankar2024-hy, Jiang2024-ia}, making autofulfillment impossible (e.g., how many rounds of questions should be asked in \cref{fig:motivation}).
At the time of writing, tasks that involve multi-step processes, decision-making, or context knowledge are likely more \textit{LLM-hard} and thus benefit more from explicit requirements. However, this is a dynamic concept that evolves as LLMs improve.

Nowadays, LLMs are widely used by \emph{end user to create reusable prompt programs}, and these prompt programs are often LLM-hard, as users frequently seek customization with various requirements, anticipating diverse inputs and expected outputs. From content creators crafting prompts for creative graphics, educators tailoring LLMs as tutors, to office workers automating workflows with GPT, people across various fields are all building natural language prompt programs.\footnote{While a prompt program can output code like traditional natural language programs \cite{jiang2022discovering}, it can also produce other outputs such as videos or GPTs applications. Note that when we refer to natural language program later in the paper, we are generally describing prompt programs.}
We argue that \emph{all these potential prompt programmers would benefit from articulating better requirements.}\footnote{In contrast, more ad-hoc interactions with LLMs, such as casual conversations with the model, may require a different set of considerations, which we consider beyond the scope of this discussion.}

\paragraphBold{The challenge of ROPE and the call for requirement-focused training}
Ideally, we hope LLM users would naturally refine requirements in prompts based on unsatisfactory model outputs.
In reality, this is challenging.
Novices struggle to \emph{understand what constitutes a requirement} in prompt engineering~\cite{Zamfirescu-Pereira2023-is}, while even experienced users struggle at \emph{extracting and expressing requirements appropriately.}
For example, users find it difficult to translate raw observations on model outputs (``I don't like how the chatbot didn't introduce itself'') to clear requirements (``Introduce yourself at the start of the conversation, and state what you can help with'')~\cite{petridis2024constitutionmaker, liu2023wants}; 
Users also cannot decide on the right level of specificity and abstraction~\cite{Liu2023-hf}, e.g., use overly general keywords where more granular, domain-specific requirements are needed for LLM-hard tasks \cite{Nguyen2024-sd, Nam2024-td}.

We hypothesize that \emph{users need explicit guidance to prioritize requirements during prompt creation}. Effective training should help users recognize the importance of clear, complete requirements and develop skills that connect abstract ideas with concrete model behaviors --- goals we address in our training design.

%% file: sections/training_design.tex
\section{Training and Evaluation Design for \sysname}
\label{sec:design-training}

We develop a \textbf{training and assessment suite} to help users improve their ability to write \emph{accurate and complete} requirements for instructing LLMs. 
We adopt a backward design method~\cite{Wiggins2005-ev}, a well-established instructional design approach that starts by identifying the desired learning outcomes --- in our case, writing effective requirements for LLMs --- then works backward to develop assessments and training aligned with goals. Aligned assessments and training are critical to ensure that what is taught directly prepares participants for the skills they are expected to demonstrate.

Through multiple pilot studies with novices and experts ($n = 10$), we refine three key components of design: 
(1) realistic tasks that mirror real-world prompting challenges (\cref{subsec:design-task});
(2) assessments that connect requirement quality to LLM outcomes (\cref{subsec:design-assessment}); and
(3) deliberate practices with feedback in a system (\cref{subsec:design-interaction}). 
Beyond training (\cref{sec:user-study-result}), our \sysname assessment and materials also enable us to build more nuanced understandings on how requirements affect LLM outputs.

\subsection{Task Design: LLM-Hard Prompt Programs for Replication}
\label{subsec:design-task}

\input{figures/table_task_type}

We aim for novices to eventually develop the ability to articulate requirements for their own prompt programs. However, to begin their training, we need to provide a set of concrete sample tasks for them to practice and for us to evaluate their progress. Below, we outline the six carefully designed tasks.

\paragraphBold{Task setup}
To ensure the training and evaluation process is both realistic and representative, our primary objective is to have users \emph{write natural language prompts to instruct LLMs in generating applications}. 
Instead of open-ended tasks --- which are difficult to assess and provide consistent feedback on --- we aim to \emph{provide clear ground truths for evaluation and training}. To this end, users are asked to write prompts to {\textbf{develop an LLM app given examples}}. 

Each task consists of two key components:
(1) A set of \emph{interaction examples} from which users can deduce requirements; and (2) A set of \emph{gold requirements} that precisely reflect the app's behavior.
Ideally, an expert skilled in requirement-driven prompt engineering should be able to express all the gold requirements within their prompts, guiding the LLM to generate an app that {produces all the interaction examples.} 

This setup requires users to derive requirements from provided examples, {which represents a realistic form of communication from a client to a developer}. It captures actual challenges of requirement articulation, especially for LLM-hard tasks in the real world (as described in \cref{sec:paradigm}): 
{Users as LLM app developers} often have access to some text inputs or visual examples like a demo or a Figma mockup, often need to refine their requirements by analyzing available outputs, but do not always know how to connect model outputs with abstract requirements. 
We discuss potential future work of open-ended tasks in \cref{subsec:limitation}.

\paragraphBold{Task selection and customization}
To cover broad types of realistic prompt programs users may write, we incorporate both GPTs applications directly powered by natural languages, as well as natural-language-to-code applications.
We prioritize tasks most suitable for targeted and engaging training, through two dimensions:
(1) we prioritize \emph{generic tasks} over specialized ones (e.g., data science tasks), to ensure that the challenge lies in specifying requirements rather than applying domain-specific expertise, preventing construct irrelevance; and (2) we opt for \emph{tasks with visual interfaces}, enabling users to interact with and visualize outputs in a manner that is accessible to novices.

With these criteria in mind, we select three GPT-driven tasks ---\task{Outline Assistant}, \task{Trip Advisor}, and \task{Email Proofreader} --- along with three coding tasks presented as visual game interfaces: \task{Connect4}, \task{Tic-Tac-Toe}, and \task{Tetris} (see examples in \cref{table:task}). 
{The GPTs tasks provide interaction examples in terms of textual chat histories, and the Game tasks provide interaction examples via an interactive visual interface.}
All these tasks are inspired by real-world applications and have reference requirements, though we make necessary adaptations to ensure they are \emph{LLM-hard} problems. 
The Game tasks are sourced from the Introduction to CS course at our institution, each with reference requirements provided by the instructor.
They require minimal customization, as instructors have already incorporated LLM-hard requirements to prevent students from cheating using large language models. 
For the GPTs tasks, we adapt real-world GPTs prompts,\footnote{Writing Assistant prompt: \href{https://github.com/linexjlin/GPTs/blob/main/prompts/Writing\%20Assistant.md}{link}, Trip Advisor prompt: \href{https://github.com/B3o/GPTS-Prompt-Collection/blob/main/17/Trip\%20Advisor.md}{link}, Email Proofreader prompt: \href{https://github.com/linexjlin/GPTs/blob/main/prompts/Email\%20Proofreader.md}{link}} and make them more LLM-hard by using frameworks like IFEval to introduce customized requirements such as format constraints \cite{Zhou2023-tb}.
We then derived the requirements by parsing the underlying prompts through expert annotation and pilot testing.\footnote{Detailed instructions of the task, as well as the requirement breakdown, are provided in supplemental materials.}

While these tasks are aligned for the training and assessment design~\cite{Wiggins2005-ev}, they span a variety of application domains with unique requirements, preventing novices from relying on memorization and promoting meaningful learning. As shown in \cref{table:task}, the GPTs and Game tasks also offer complementary coverage on requirement specificity and modality.

\paragraphBold{Task validation} 
We distribute the six tasks in assessments and the training session. We use \task{Tetris} and \task{Email Proofreader} in training, and the rest for pre- and post- assessments before and after training in a counterbalanced design.
In pilot studies, we confirmed that (1) we could distinguish prompting novices and experts using the tasks, with the average novice performance being 28\% and expert achieving full scores, (2) all tasks could be implemented by LLMs, with the average highest score being 91\%, and (3) the assessment tasks were all comparably challenging for users.

\subsection{Assessment Design: Requirement-Focused Intrinsic and Extrinsic Evaluation}
\label{subsec:design-assessment}

To evaluate users' ability to instruct LLMs effectively, we assess both the quality of the user prompts (requirements within the prompt) and the prompt's impact on the quality of the LLM's output. Thus, for each user task completion, we calculate its \emph{Overall Score} as the average of two components, \emph{Requirement Quality} and \emph{LLM Output Quality}:

\begin{itemize}[labelwidth=*,leftmargin=1.3em,align=left]
    \item \emph{Requirement Quality}: 
    This intrinsic metric assesses, ``Does the user's prompt accurately and comprehensively cover all requirements?'' It measures the correctness and completeness of the requirements by comparing those extracted from the user's prompt to expert-defined reference requirements for each task (described in \ref{subsec:design-task}).
    We operationalize requirement quality as the percentage of reference requirements that are correctly described in the user's prompt. Drawing from the requirements defect taxonomy~\cite{Alshazly2014-qr, Montgomery2022-vq}, we mark both \emph{commission errors} (incorrect, inconsistent, or ambiguous requirements) and \emph{omission errors} (missing reference requirements) in users' prompt, so a requirement is only correctly described if it is free from commission and omission defects. 

    \item \emph{LLM Output Quality}: 
    This extrinsic metric evaluates, ``Can the user's prompt successfully guide the LLM to achieve the intended goals?'' It measures the proportion of desired features implemented in the LLM-generated output that align with the reference (\cref{subsec:design-task}). 
    We pass users' prompts to GPT-4o and compare the generated output to the reference program.\footnote{The output is generated using a prompt similar to \texttt{``Follow this prompt: \{user\_prompt\}''}. Further details are in the supplemental materials.} 
    Since many features are difficult to evaluate automatically (e.g., GPTs behaviors cannot always be checked via rules), we rely on manual evaluation.
    For games, we interact with the generated program to verify whether the requirement features are correctly implemented. For GPT interactions, we grade whether the GPTs' replies display expected behaviors (e.g., asking a follow-up question when the user's input is ambiguous). We grade 10 complete conversations per GPTs.
\end{itemize}

\paragraphBold{Assessment validation}
Three of the authors discuss and iterate the grading rubrics on the tasks during the pilot study. 
We confirm that an expert can measure requirement quality using the percentage of correct requirement clauses in novice prompts, and that this correlates positively with the expert's judgment of overall requirement quality (Spearman's $\rho = 0.66$).
For grading the \emph{Overall Score} in the user study (\cref{sec:user-study-design}), two authors (one of whom did not participate in the development of the rubrics during the pilot) independently grade 10\% of the randomly chosen pre-post test responses. 
We check the inter-rater reliability between the authors' \emph{Overall Scores} by calculating the Intraclass Correlation Coefficient (ICC) \cite{Koo2016-lv}, finding strong reliability ($ICC = 0.9$, \text{95\% Confidence Interval } = [0.7, 0.98]). Any discrepancies in scoring are resolved through discussion, after which the authors individually grade half of the participants' responses.
Detailed rubrics are provided in supplemental materials.

\begin{figure*}
    \centering
    \includegraphics[trim={0 0cm 0cm 0cm}, clip, width=0.9\linewidth]{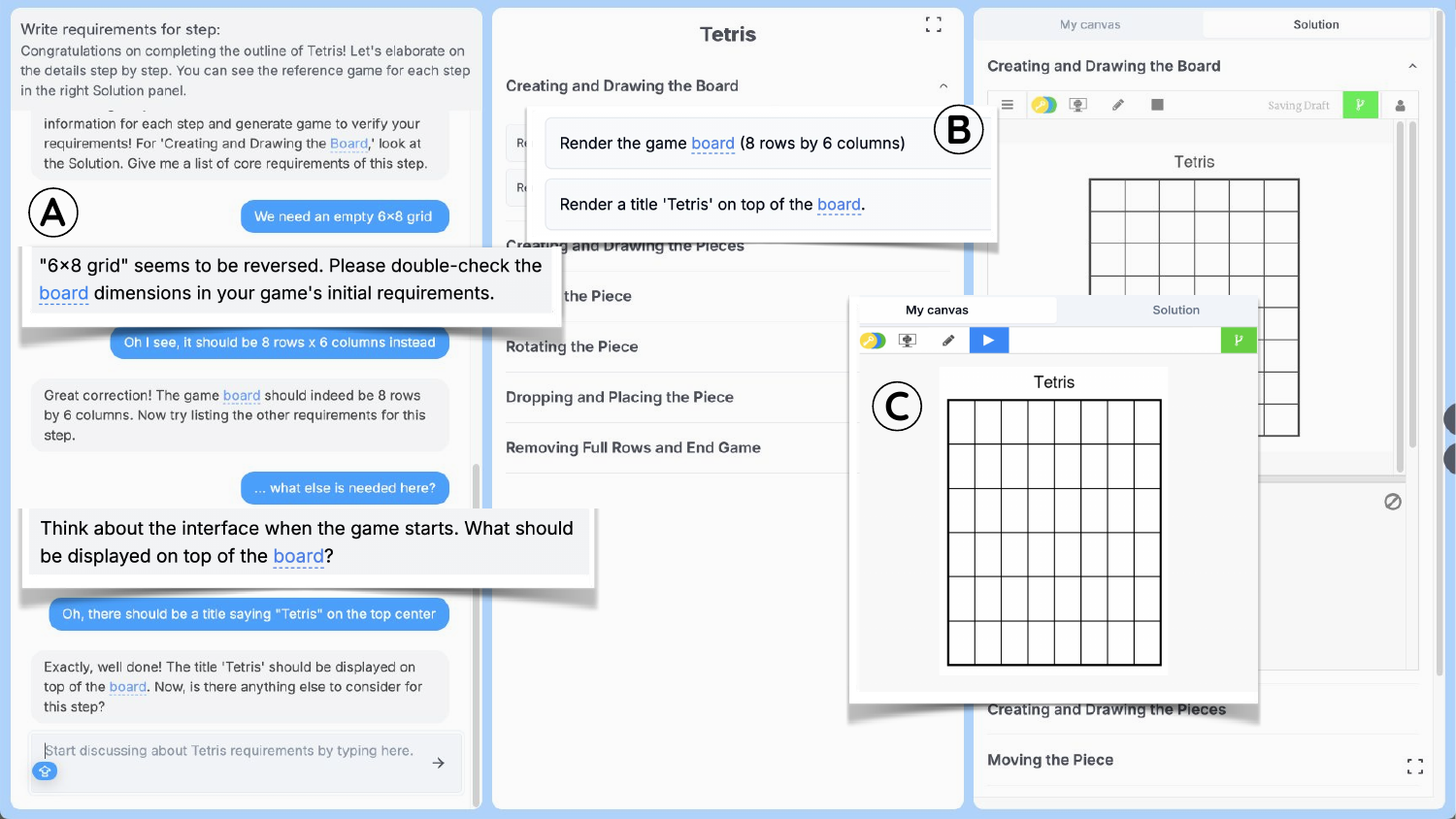}
    \caption{Our \sysname training interface, with three types of feedback on requirement defects: (A) \emph{Chat-based hints} on incomplete or inaccurate requirements, (B) \emph{Reference requirement examples} to reinforce appropriately identified and expressed requirements, and (C) \emph{LLM output counterfactual} to support user's reflection on incorrect or ambiguous requirements.}
    \label{fig:feedback}
    \Description{Screenshots of the ROPE training interface showing feedback types: (A) Chat-based hints for requirement defects, (B) Reference requirement examples, and (C) LLM output counterfactual to reflect on incorrect or ambiguous inputs. The scenario focuses on developing Tetris game requirements.}
\end{figure*}

\subsection{Interactive Training Mechanism: Dedicated Practice and Feedback on Requirement Defects}
\label{subsec:design-interaction}

We design a training mechanism to support deliberate practice on requirement elicitation and refinement by \emph{disentangling requirement articulation from peripheral tasks like persona crafting or paraphrasing in existing prompt engineering instructions.} We apply key learning principles like scaffolding and worked example~\cite{Koedinger2013-bp}, and we implement the training into an interactive interface as shown in \cref{fig:feedback}.

In this training, a user (a prompt novice) is asked to develop one Game (\task{Tetris}) and one GPTs (\task{Email Proofreader}) {given the interaction examples}, \emph{solely by describing the requirements of the app as accurate and complete as possible}. 
For example, to develop the customized \task{Tetris} game in \cref{fig:feedback}, the user will first start by outlining the main milestones (e.g., creating the game board, handling piece placement, etc.), and then provide more detailed specification per step (e.g., define the size of the game board).
Novices are not expected to achieve full success on their first try; we use three types of \textbf{requirement-focused feedback} to guide users in continuously refining their requirements:

\begin{itemize}[labelwidth=*,leftmargin=1.3em,align=left]
    \item \emph{Conversational hint and clarification}, via chatbot (\cref{fig:feedback}A):
    We create a tutor chatbot to provide feedback on users' requirement mistakes. For instance, the chatbot may ask \quoteinline{What's on top of the board?}, when a user \textit{misses} the requirement for \req{Render a title `Tetris' on top of the board} (\emph{omission error}). As shown in \cref{fig:feedback}A, the chatbot may suggest the user to \quoteinline{double-check the board dimensions} if the user \textit{incorrectly} reverses the number of rows and columns for the requirement \req{Render the game board (8 rows by 6 columns)} (\emph{commission error}). The textual feedback encourages critical thinking on \emph{missing} or \emph{incorrect} requirements and offers a natural, conversational experience to discuss and reflect on requirements.
    
    \item \emph{Reference requirement example}, via requirement working document (\cref{fig:feedback}B): 
    We progressively reveal reference requirements as feedback when users correctly identify them during interaction with the chatbot. The expert-written reference examples are provided to \emph{reinforce correct requirements}, helping users understand how to formalize and organize requirements. 
    
    \item \emph{LLM output counterfactual}, via generated visualization (\cref{fig:feedback}C):
    If novices produce incorrect requirements that can be visually demonstrated (e.g., wrong board dimensions), we provide visual feedback by generating flawed programs that implement the \emph{incorrect} requirements, or maliciously misinterpret \emph{ambiguous} requirements. For example, a vague requirement like \quoteinline{Use keys to move pieces} might result in a flawed Tetris game where pieces can move upward, exposing the unspecified allowed movement in requirement. 
    The visual counterfactual is currently limited to the controllable game code generation; as GPTs outputs tend to be less predictable, we display static chat histories as illustrative examples for GPTs.
\end{itemize}

Note that our assessment tasks require users to write complete prompts similar to those in \cref{fig:motivation}, we deliberately avoided having users write prompts during training. Instead, users engage in conversational interactions designed to continuously encourage them to think about requirements. Additionally, users' written responses do not directly trigger LLM output generations. We use reference requirements to guide the LLM generation, ensuring that all feedback reflects the quality of the requirements alone. This approach ensures that novices remain focused on improving their requirement articulation skills, without distractions from other factors.

\paragraphBold{Interaction and feedback implementation}
We use OpenAI GPT-4o to power the interface,\footnote{The state-of-the-art LLM at the time of writing: \url{https://openai.com/index/hello-gpt-4o/}. We used a temperature of 0.3 for code generation and 0.7 for other tasks. Refer to supplemental materials for our prompts.} leveraging its interactive nature to adaptively provide feedback on the diverse set of possible user inputs.
However, feedback generation does not rely on the model's inherent reasoning capabilities but is anchored in our predefined reference requirements.
Specifically, we always have GPT-4o to compare the current user requirements to the reference, and select the most critical defect to provide feedback on (roughly, mismatched requirements are considered most critical, then missing requirements, then ambiguous ones).
Then, we have the LLM respond targetedly to the defect.
For example, visual counterfactuals on game tasks are generated by having the LLM minimally edit reference Python code based on the identified incorrect requirement.
To enhance interaction, we also implement features like highlighting special variables (e.g., board, keystrokes) in conversations and as hyperlinks for easy cross-referencing across the interface.

\paragraphBold{Validation on feedback generation}
\label{subsec:feedback-val}
We analyze the interface log data to evaluate the feedback quality for all types of feedback (\cref{fig:feedback}A, B, C).
From our user study (\cref{sec:user-study-design}), we collected a total of 635 interaction turns from \task{Tetris} (ranging from 11 to 73 turns per user) and 180 turns from \task{Email Proofreader} (ranging from 4 to 20 turns per user).
From this, we randomly selected 10 different users and annotated conversations for 5 users per task, resulting in 151 \task{Tetris} turns, 66 \task{Email Proofreader} turns, and a total of 113 annotations per feedback type.
For each LLM turn, we assess: (1) Is the feedback needed? (2) Is the feedback provided? (3) Is the provided feedback correct? Given the answers to the three questions, each feedback was categorized as correct, incorrect, irrelevant, missing, or not provided (\cref{tab:metrics}). 
For example, \textit{irrelevant feedback} are those that are (1) not needed but (2) provided. Note that the right feedback includes both \textit{correct feedback} that was (1) needed, (2) provided, and (3) correct, and \textit{not provided feedback} that was (1) not needed and (2) not provided.

\input{figures/table_llm_eval}

Two authors independently annotated one participant's chat log for each task, and we measured Inter-Rater Reliability (IRR) by calculating Krippendorff's $\alpha$ \cite{castro-2017-fast-krippendorff}, achieving a high agreement rate ($\alpha = 0.87$) across all feedback types (see \cref{tab:metrics} for the detailed breakdown). Any discrepancies were discussed and resolved, after which one author completed the remaining annotations. The final analysis revealed a correct feedback rate of 88.6\%.

%% file: figures/table_task_type.tex
\begin{table*}[]
\centering
\caption{An overview and comparison on the task types in our training.}
\footnotesize
\begin{tabular}{p{0.09\textwidth}  | p{0.41\textwidth}|p{0.41\textwidth}}
\toprule
& \textbf{GPTs Task} & \textbf{Game Task} \\
\midrule
Task Type 
    & Customized LLM powered directly by natural language
    & Natural Language to code (from CS assignment) \\
    
\midrule
Example \& \newline Requirements \emph{(Rubrics)}
    & \raisebox{-0.95\height}{\includegraphics[trim={0cm 7cm 15cm 0cm}, width=\linewidth]{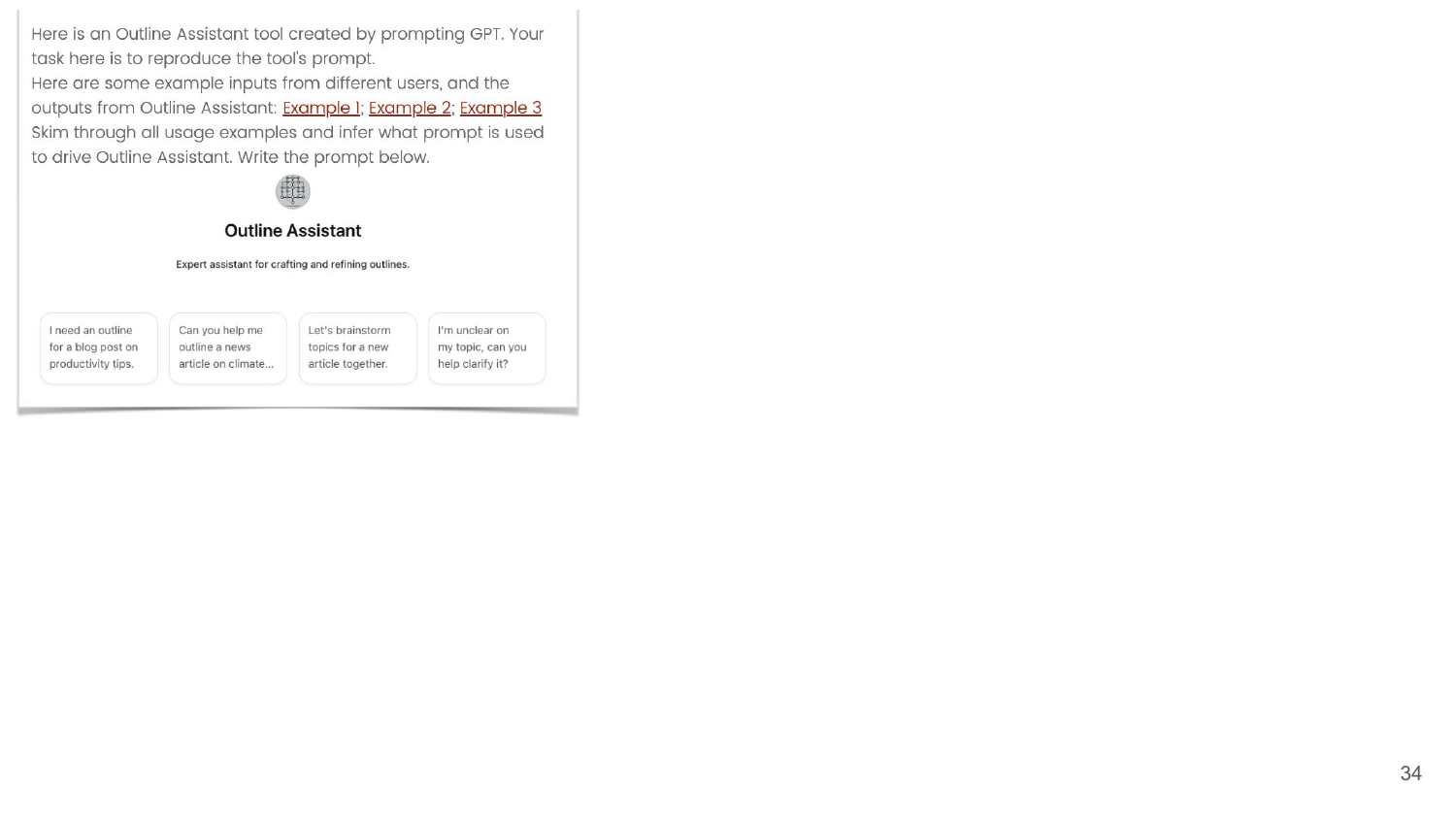} }
    & \raisebox{-0.95\height}{\includegraphics[trim={0cm 7cm 15cm 0cm}, width=\linewidth]{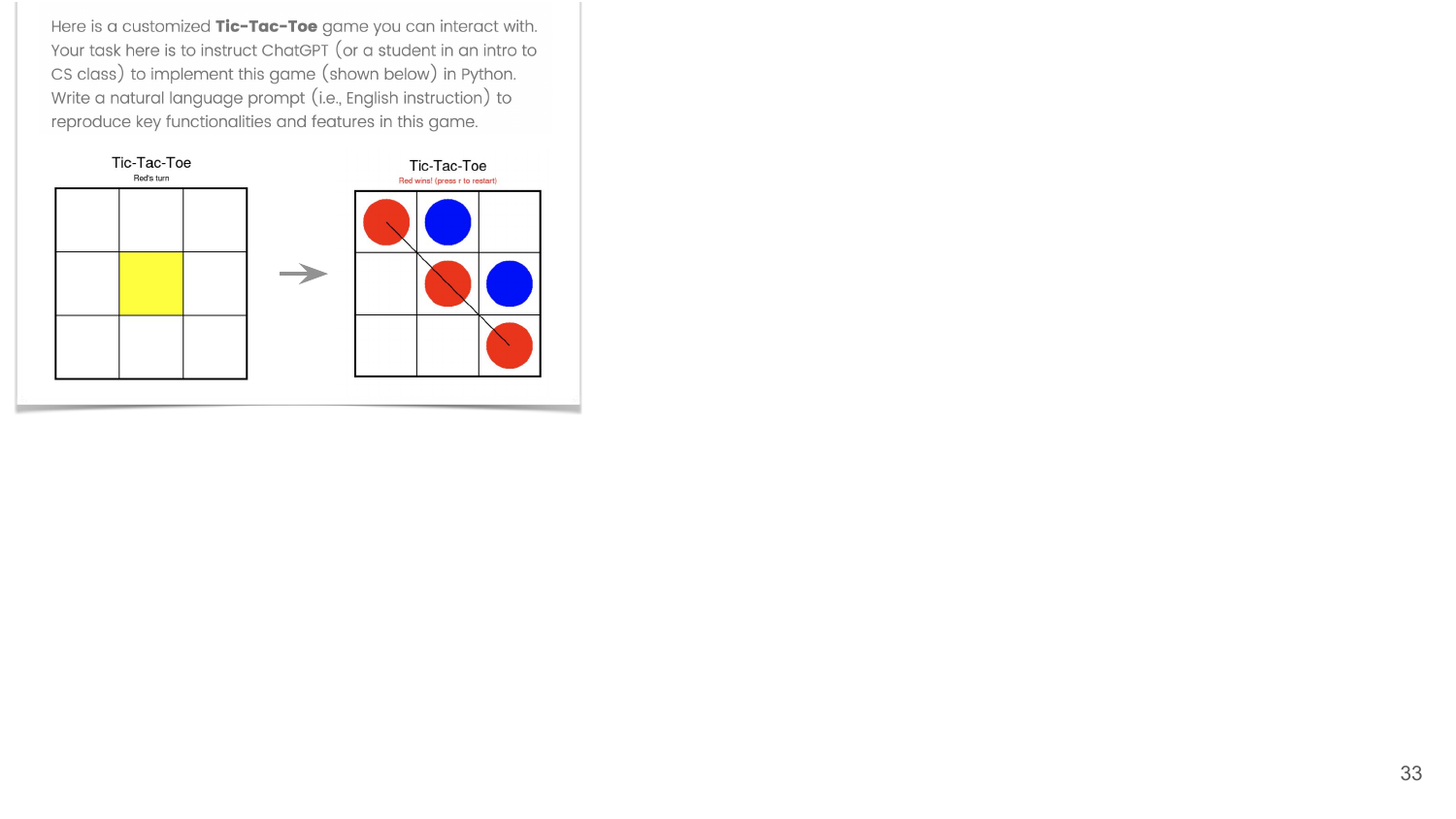}} \\

     & 1. Use follow-up questions to let the users clarify and specify their needs, e.g., target audience, length, focus, tone, and style. \newline
        2. Offer advice for potential modification direction. \newline
        {\color{cquote}(6 in total)} 
    & 1. Render the title Tic-Tac-Toe on top of the board. \newline
        2. Display the corresponding message (e.g. Red's turn) \newline
        3. Keypress 'r' to restart the game \& reinitialize the board. \newline
        {\color{cquote}(9 in total)} \\

\midrule
Specificity
    & Higher-level requirements covering diverse categories, but can have subjective interpretations 
    & Lower level requirements akin to implementation specification but more verifiable \\
\midrule
Modality 
    & Convey requirements through textual examples --- users are given 2-3 chat histories to control for task time and interaction exposure. It is representative of \emph{status-quo} prompt iteration. 
    & Convey requirements visually --- users are given an interactive game. The interaction reduces ambiguity that is common with few examples and requires less English reading skills. \\ 

\bottomrule
\end{tabular}
\label{table:task}
\end{table*}

%% file: figures/table_llm_eval.tex
\begin{table*}[ht!]
\small
\centering
\caption{Proportions of feedback types across different feedback sources and IRR (Krippendorff's $\alpha$) of human annotations}
\vspace{-10pt}
\label{tab:metrics}
\Description{Description of the table.}
\begin{tabular}{l | ccc | cccc | c}
\toprule
\multirow{2}{*}{\textbf{Feedback Source}} 
    & \multicolumn{3}{c|}{\textbf{Right Feedback}} 
    & \multicolumn{4}{c|}{\textbf{Wrong Feedback}} 
    & \multirow{2}{*}{\textbf{IRR ($\alpha$)}} \\ 
 & \textbf{Correct} & \textbf{Not Provided} & \textbf{Total} 
 & \textbf{Incorrect} & \textbf{Irrelevant} & \textbf{Missing} & \textbf{Total}  \\ \midrule
\textbf{Chat} (e.g., Fig. \ref{fig:feedback}A) 
& 69.03\% & 20.35\% & 89.38\% & 10.62\%  & 0.00\% & 0.00\% & 10.62\% & 0.86 \\ 

\textbf{Ref Example} (Fig. \ref{fig:feedback}B)
& 48.67\% & 38.05\% & 86.73\% & 1.77\%  & 5.31\% & 6.19\% & 13.27\% & 0.88 \\

\textbf{LLM Output} (Fig. \ref{fig:feedback}C)
& 3.75\%  & 86.25\% & 90.00\% & 0.00\%  & 5.00\% & 5.00\% & 10.00\% & 0.68 \\

\midrule
\textbf{All Feedback} 
& 44.44\% & 44.12\% & 88.56\% & 4.58\%  & 3.27\% & 3.59\% & 11.44\% & 0.87 \\

\bottomrule
\end{tabular}
\end{table*}

%% file: sections/user_study.tex
\section{User Study Design}
\label{sec:user-study-design}

We conducted a user study to understand whether our requirement-focused training is effective in improving novices on writing requirements in prompts, and whether requirement-focused training is \emph{more} effective than \emph{status-quo} prompt engineering training.
We also examined how automatic prompt optimization affects performance.

\paragraphBold{Study procedure}
To capture the requirement-focused training gains, we utilized a \textbf{pre-test and post-test approach}, a standard instruction evaluation method~\cite{Rogaten2019-ii}, comparing participants' performance on isomorphic tasks before and after \sysname training (within-subject). 
To compare requirement-focused training with standard training, we utilized a \textbf{randomized-control design} (between-subject). We randomly assigned participants to either the experimental condition (\sysgroup group), where they receive requirement-focused training in our interface, or the \emph{control} condition ({\ctrlgroup} group), where they receive a standard prompt engineering tutorial and then self-practice using ChatGPT. 

Our user study --- for either the \sysgroup or the \ctrlgroup group --- is 1.5 hours long. 
In the study, participants first completed a two-minute \emph{pre-survey}, providing demographic information and rating their prompting experiences (e.g., confidence and self-perceived productivity using ChatGPT) with 7-level Likert Scale questions. 
Then, they started a \textit{pre-test} (20 minutes) to write prompts to replicate key features on existing programs (\cref{table:task}) --- one Game (\task{Tic-Tac-Toe} or \task{Connect4}) and one GPTs (\task{Outline Assistant} or \task{Trip Advisor}).
They proceeded to the 40-minute \textit{training session} (different interfaces depending on experiment group) on two tasks (\task{Tetris} and \task{Email Proofreader}), followed by a \textit{post-test} that contained different Game and GPTs tasks (20 minutes).\footnote{
The pre- and post- assessments each contained one Game task and one GPTs task (\cref{subsec:design-task}), counterbalanced to mitigate problem sequencing bias or different task difficulties. 
We found that the two versions of the tests have comparable difficulties from pre-test results (an average overall score of 21.0\% and 25.6\% for each version).}
To further capture participants' iterative prompting behaviors, for the post-test GPTs task, participants were also asked to feed their prompt into ChatGPT and iterate their original prompt by observing the ChatGPT outputs.
They completed the study with another two-minute \textit{post-survey}, providing feedback with open-ended or Likert Scale questions on their experience and perceptions during the training. 
Participants were compensated with a \$20 Amazon Gift Card. Please refer to supplemental materials for our study materials.

\paragraphBold{Control group (\ctrlgroup) design}
We designed the \ctrlgroup group to replicate the ``business-as-usual'' scenario, where novices learn prompt engineering through publicly available resources and self-practice.
Specifically, \ctrlgroup group participants first watched a 20-minute YouTube tutorial,\footnote{Relevant sections from \href{https://www.youtube.com/watch?v=_ZvnD73m40o}{Prompt Engineering Tutorial – Master ChatGPT and LLM Responses}}
covering best prompting practices including clear instructions, adopting personas, specifying format, limiting the scope, and few-shot prompting. The tutorial included various prompting examples such as poem writing, code-based data processing, essay summarizing, etc.
Afterward, participants self-practiced writing prompts for the two training tasks, iterating with ChatGPT directly. For the game task, we offered participants an online code compiler so participants could see games rendered in real-time, similar to the visual feedback in our interface (\cref{fig:feedback}C).
Similar to the \sysgroup group setup, our \ctrlgroup group offers practices on coding and GPTs tasks to establish task familiarity. However, unlike our requirement-focused training where users practice writing only requirements without actually writing prompts, the conventional practice for \ctrlgroup group directly train participants on writing prompts, which is closer to what users actually write in pre-post tests, and the direct interaction with ChatGPT might help users build a better mental model on LLM behaviors.

\paragraphBold{Data collection}
To understand participants' perceived experiences, prompting behaviors, and learning outcomes, we collected participants' prompt responses, survey answers, time on task, as well as interaction log data during the training phase. Afterward, we collected the LLM outputs by feeding participant prompts into OpenAI \texttt{GPT-4o}.\footnote{We used a temperature of {0} for \texttt{GPT-4o} to generate deterministic outputs, please refer to our supplemental materials for more details.}
Using the evaluation method and rubrics we developed for each of the assessment tasks \cref{subsec:design-assessment}, we graded the pre- and post- tests using users' \textit{original prompts} for all tasks from the test, on a scale of 0\% to 100\%. We calculated the \emph{Overall Scores} as the average of the \emph{Requirement Quality} and \emph{LLM Output Quality} scores.
To further understand the effect of model advancement, we collected LLM outputs with the newer, more powerful reasoning model OpenAI \texttt{o3-mini}\footnote{\url{https://openai.com/index/openai-o3-mini/} with the default settings.} and conducted the same analysis. The results and comparisons against \texttt{GPT-4o} are presented in \cref{subsec:res-o3mini}.

To understand the effect of an optimizer on a prompt (as discussed in \cref{subsec:PO}), we further use the Prompt Maker,\footnote{\url{https://chatgpt.com/g/g-hhh4w3eov-prompt-maker}: from a simple prompt to an optimized prompt.} a popular optimizer with 50k usages in GPT Store, to refine participants' \emph{original prompts} into \emph{optimized prompts}. 
This optimizer is easily applicable to any kind of prompts (\cref{fig:motivation} shows an example of its output), though it is a rather preliminary version, as it rephrases user prompts in a rule-based manner without criticizing the LLM outputs of the original prompt.
We evaluated the optimized prompts in the same way as the human prompts (\cref{subsec:design-assessment}), including generating and grading corresponding LLM outputs.

\paragraphBold{Participants}

We recruited users with limited or no NLP background from different institutions in the United States. 
We conducted studies with 32 participants randomly assigned to the \sysgroup or \ctrlgroup group, and we removed two participants {who disengaged during the study} from the analysis. In total, we have 30 participants (S1-30, $n=15$ for each condition) --- 19 female, 10 male, 1 agender, and 18 non-native English speakers (9 in each condition), with an average age of 26.
Our participants have a diverse range of backgrounds including HCI, education technology, communication, graphic design, mechanical engineering, information systems, literature, psychology, and economics. We asked the participants to rate their familiarity with LLMs and complex prompts in the pre-survey, and the average rating is 3 out of 7.

%% file: sections/user_study_result.tex
\section{User Study Results}
\label{sec:user-study-result}

\subsection{Learning Gain: Requirement-focused Training Helps Students Instruct LLMs}

We start by answering our central research questions: \emph{Can we help end-users better instruct LLMs through requirement-focused training?}
We quantitatively measured learning gains by capturing participant performances from the pre-test to post-test within each experiment group, and evaluated the effectiveness of the two training approaches by comparing the \sysgroup and \ctrlgroup group.
We further unpacked the quantitative results through analyses of participants' Likert Scale ratings,\footnote{We investigated whether the learning gains are associated with any demographic factors like age and did not find any significant indicator. There is a weak negative correlation between self-perceived familiarity with LLM and learning gains, which means that the lower LLM familiarity the participants rated, the more they learn from the training ($\rho = -0.2$).} as well as the transcribed comments during their training sessions and in post surveys.

\paragraphBold{Requirement-focused training is effective}

\begin{figure}[t!]
    \vspace{-10pt}
    \includegraphics[width=1\linewidth, trim={0 2.5cm 9cm 0cm},clip]{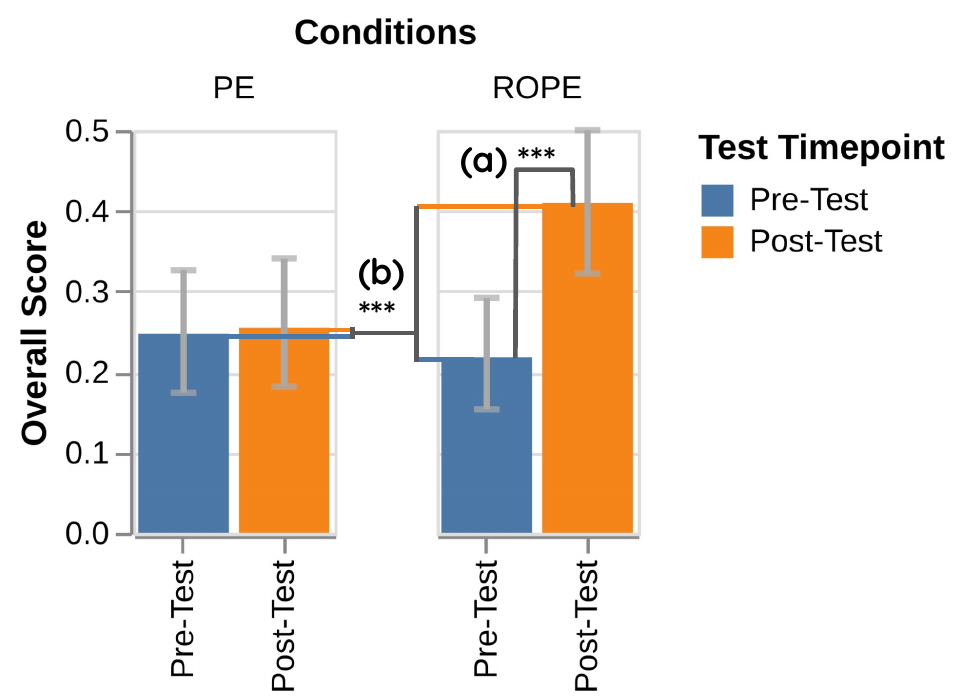}
    \caption{The pre-test and post-test overall scores between \ctrlgroup and \sysgroup conditions. (a) Overall scores for \sysgroup significantly improved from pre- to post-test; (b) \sysgroup achieved significantly higher learning gains (post $-$ pre) than \ctrlgroup (*** denotes $p < 0. 001$).}
    \label{fig:prepost-gains}
    \Description{Bar chart comparing pre-test and post-test scores between PE (Prompt Engineering) and ROPE conditions. ROPE shows a statistically significant (p < 0.001) improvement in overall scores from pre- to post-test, unlike PE.}
    \vspace{-5pt}
\end{figure}

Using a two-tailed paired t-test,\footnote{We did a linear mixed effects regression to account for the correlation across participants and task (both as random effects) of the repeated measures for the pre- and post-test times and for the requirement and LLM output scores. We found the same pattern of significant results as revealed by the t-tests.} we found that the overall scores for participants in the \sysgroup group significantly improved by 19.1\%\footnote{\emph{Requirement Quality} score improved by 25.4\% and \emph{LLM Output Quality} score improved by 12.7\%, both significant ($p < 0.05$). Pre-test performances were not correlated with learning gains. } from pre- to post-test ($p < 0.001$, from $21.9\% \pm 14.6\%$ to $40.9\% \pm 18.4\%$), an almost two-fold increase (as shown in \cref{fig:prepost-gains}a). This demonstrated that \emph{requirement-focused training enabled participants to instruct LLMs more effectively,} and we achieved our desired goal: \emph{to concentrate users' attention on only requirement iterations during the training, but make sure users can still apply their learning in overall prompt writing.}
Analyzing requirement defects in \sysgroup participants' prompts, we observed a noticeable decrease in omission errors (from 5.6 to 3.2 per participant), but a slight increase in commission errors (from 0.5 to 0.7 per participant). This indicated that participants \emph{became more aware of requirements, but need additional training for requirements clarity and accuracy.}

\sysgroup participants' self-reflection highlighted that they understand the value of \emph{emphasizing and iterating} on clear and complete requirements, which may contribute to their improvement. 
For example, 7 out of 15 \ctrlgroup group and 13 out of 15 \sysgroup group participants mentioned ``be specific'' in their answers to prompting strategies in post-survey. When asked about what they learned during the instruction, four \ctrlgroup participants noted to ``give exact instruction,'' as described by S23, \quoteinline{we should give as much information as we can to ChatGPT}. Meanwhile, some \sysgroup participants developed a more in-depth understanding of articulating requirements in prompts. For example, six \sysgroup participants explicitly noted the connection between requirements and prompts.
S30 described requirements as a way \quoteinline{to be more clear and not confuse the system by giving too much unnecessary information}, and S8 noted that a requirement-focused approach helped them \emph{organize their thoughts}: 
\quoteinline{I learned to organize my requirements logically so we can easily revise and improve them}. 
S10 neatly described their shift towards requirement-focused prompting strategy: \quoteinline{sometimes when I write prompts, [I find] steps are hard to be clearly divided, or I didn't consider to divide them that detailed. However, it's important to do so, as it appears to give LLM more direct and clear instruction. When the steps are divided, it's easier to see the missing details in my original prompts too}.

\paragraphBold{Requirement-focused training is more effective than standard prompt engineering practice}
From the post-test prompt data, we observed that requirement-focused training and standard practice had distinct effects on how participants wrote prompts. \sysgroup group spent more time (19 minutes vs. 14 minutes) and wrote longer prompts (796 vs. 458 characters) than the \ctrlgroup group. In 100 randomly selected pairwise comparisons, \sysgroup participants produced more structured prompts 87\% of the time, consistent with their self reflections reported above.

These prompts also revealed learning gain differences.
A two-sample t-test showed that the \sysgroup achieved significantly higher learning gains compared to the \ctrlgroup group ($p < 0.001$), as shown in \cref{fig:prepost-gains}b. 
In fact, while \sysgroup significantly improved (19.1\% as mentioned above), we did not observe much gain in \ctrlgroup group's pre-to-post score (0.7\%).
The contrast suggested that \emph{novices indeed could not acquire requirement articulation skills through standard prompting training alone.}

Participant feedback supported our hypothesis that \emph{requirement articulation skills do not naturally emerge}. 
While we hoped interacting with LLMs would help participants build mental models of LLM behaviors, \ctrlgroup participants found the LLM response too unpredictable for effective self-practice.
Similar to prior work reporting the challenges of prompting nondeterministic LLM~\cite{Zamfirescu-Pereira2023-is, Nguyen2024-sd}, many got frustrated during their unfruitful self-practice sessions: \quoteinline{I was confident with my communication ability at first but later during the tasks I was frustrated with my communication skill with AI} (S11).
The unpredictability distracted participants from applying the best practices taught in the tutorial, as S19 noted:  \quoteinline{I lost some of the concepts at the end that looked very minor in determining the effectiveness of the prompt}.

\paragraphBold{Requirement-focused training encouraged more targeted and iterative prompting}
The aforementioned pre- and post-tests analyses revealed participants' learning gains in \emph{writing initial prompts for natural language programs}.
Beyond initial prompts, we analyzed participants' prompt iteration behaviors on the GPTs tasks at the end of post-tests. We observed that \emph{requirement-focused training led users' to make more requirement-related changes during iterations}.

In the \ctrlgroup group, most participants either chose not to make any iterations (5 out of 15) or only made superficial edits (e.g., altering wording or grammar, 8 out of 15), with only 2 adjusting prompt requirements.
In contrast, after \sysname training, participants \emph{were more likely to engage in substantive requirement engineering}, updating or adding specific requirements instead of making random edits typical of end-user prompt engineering.
Eight of 15 \sysgroup participants made meaningful requirement adjustments (5 were fully successful, 2 made partial progress, and 1 introduced an incorrect requirement).
Participants were also aware of their requirement-centric iterations: \quoteinline{[My strategies were to] break down my requirement into several key points, use examples, iterate, self-check if more details are needed, or if more steps should be elaborated, ..., see the test result using examples} (S10).
Overall, \sysname training effectively equipped users with a more structured and systematic approach to prompt generation.

\paragraphBold{Key to success: deterministic feedback as a jump start}
Looking at both learning gains and behavioral changes, we speculate that the success of requirement-focused training stemmed from our use of \emph{requirement-focused feedback}.

Requirement articulation was challenging for all participants, as shown by the low pre-test scores (21.9\% for the \sysgroup and 24.8\% for the \ctrlgroup).
This poor starting point likely hindered the \ctrlgroup's ability to receive useful feedback during self-practice, as their requirements were too weak for the LLM to generate ``good enough'' model output with clear indicators on what to improve.
Consequently, \ctrlgroup participants were trapped and frustrated by LLM unpredictability --- S6 noted that \quoteinline{ChatGPT is a very malleable tool and can change responses pretty drastically depending on the prompt}, and S7 commented that \quoteinline{if ChatGPT doesn't get it, I might not be patient}.

In contrast,  our feedback loop for \sysname (\cref{subsec:design-interaction}) provided tightly controlled \emph{feedback} and \emph{program output} based on the requirements users specified, regardless of grammatical issues or natural language variances.
S8 highlighted that: \quoteinline{It's easy to understand (very logical) and can see how the changes and revises influence the system immediately}.
This deterministic feedback reinforced \sysgroup participants' belief in the importance of clear requirements, leading to them writing better initial prompts with more requirements in the post-test.
We hypothesize that these better initial prompts might also become \emph{more suitable for further iteration}. Although still imperfect, \sysgroup participants' prompts might allow the LLM to generate partially correct outputs that contain enough signal towards future improvements.

We suspect that effective prompt engineering training does require deterministic scaffolding. Novices may need reinforcement outside of LLM feedback until they reach a level where interpreting LLM responses becomes valuable. 

\subsection{In-depth Analysis: The Validity of \sysname Paradigm}

Going beyond learning gains, we dive deeper into the connections between requirements in prompts and LLM output quality, the role of optimizers, as well as the differences brought by model advancement.

\paragraphBold{Requirement quality vs. LLM output quality: promising correlation with nuances}\label{subsec:res-corr}
To examine whether more correct and complete requirements led to better LLM outputs, we calculated the correlation between the two components of our \textit{Overall Scores}: \emph{Requirement Quality} and \emph{LLM Output Quality} scores using two LLMs generating the outputs --- \texttt{GPT-4o} as in our main experiment, and \texttt{o3-mini} as explained below. 
We found a strong positive correlation, with a Spearman's correlation coefficient of {$\rho = 0.71$} \cite{schober2018correlation}.
However, task-specific analysis revealed nuances. As shown in \cref{tab:corr-by-task}, while most tasks achieved {$\rho \geq 0.7$}, \task{Connect4} showed no correlation when LLM output is generated using  \texttt{GPT-4o}. 
Upon further inspection, we found that interestingly, \texttt{GPT-4o} tends to produce two implementation types for \task{Connect4}: command line and pygame. Prompts that include interactive or visual requirements such as \req{Highlight the hovered column} often led to pygame implementations. 
If we split the \task{Connect4} correlations by the LLM output implementation type, we get reasonable numbers for both cases: $\rho = 0.89$ (command line) and $\rho = 0.58$ (pygame), suggesting that implementations may respond to requirements differently. 
For example, the requirement \req{Display player's turn message} was always implemented in command line \task{Connect4}, even if omitted in prompts. However, only 1 pygame \task{Connect4} implemented the same requirement, out of 6 participants who correctly described it. This suggests that requirements can be interdependent, with \emph{LLM-hardness} varying by implementations.

\input{figures/table_correlations}

We further analyzed how different types of requirement defects impact LLM output quality.
We found that the number of omission errors (incomplete requirements) had a stronger negative impact on \emph{LLM Output Quality Score} ($\rho=-0.49$) compared to commission errors (inaccurate requirements) ($\rho = 0.05$). This suggests that while LLMs can correct inaccuracies in requirements, they struggle to fill in missing information --- further highlighting \emph{the importance of training humans to express all their necessary requirements.}

Our findings complement existing instruction following research \cite{Zhou2023-tb,Jiang2024-ia}, which tend to rely on synthetic prompts with perfectly written requirements of a fixed set of common constraint types (\cref{subsec:PO}). Our results highlight the need for further investigation into how imperfect user requirements interact with inherent LLM biases.

\paragraphBold{Optimizer improves prompts but introduces biases too}
\label{subsec:res-optimizer}
To assess how the optimizer affected prompt quality, we experimented on a single optimizer to explore the feasibility, without generalizing to all optimizers (\cref{subsec:limitation}).

We first noticed that the Prompt Maker optimizer effectively enhanced user prompts by adding elements such as role-playing and chain-of-thought, making prompts more structured and fluent, as shown in \cref{fig:motivation}.
Besides formatting, the optimizer also made direct modifications to requirements; on average, it added 0.5 missing requirements and corrected 0.3 inaccurate requirements, reducing omission errors from $4.4$ to $3.9$ and commission errors from $0.6$ to $0.3$ per participant across all prompts. 
These changes significantly improved overall scores, as shown by a paired t-test for all users' prompts before and after optimization ($p < 0.001$).

It seemed like there existed some requirements that were \emph{LLM-hard} but not \emph{optimizer-hard}. For example,  \texttt{GPT-4o} would not implement \req{Display messages for a tie game} unless explicitly prompted, but the optimizer would automatically add this requirement to user prompts.
However, highly customized requirements, such as \req{Cross out the winning cells}, were never automatically added (0 out of 51) without explicit user input. 

\begin{figure}[t!]
    \vspace{-10pt}
    \includegraphics[width=.95\linewidth, trim={0 2.5cm 8.5cm 0cm},clip]{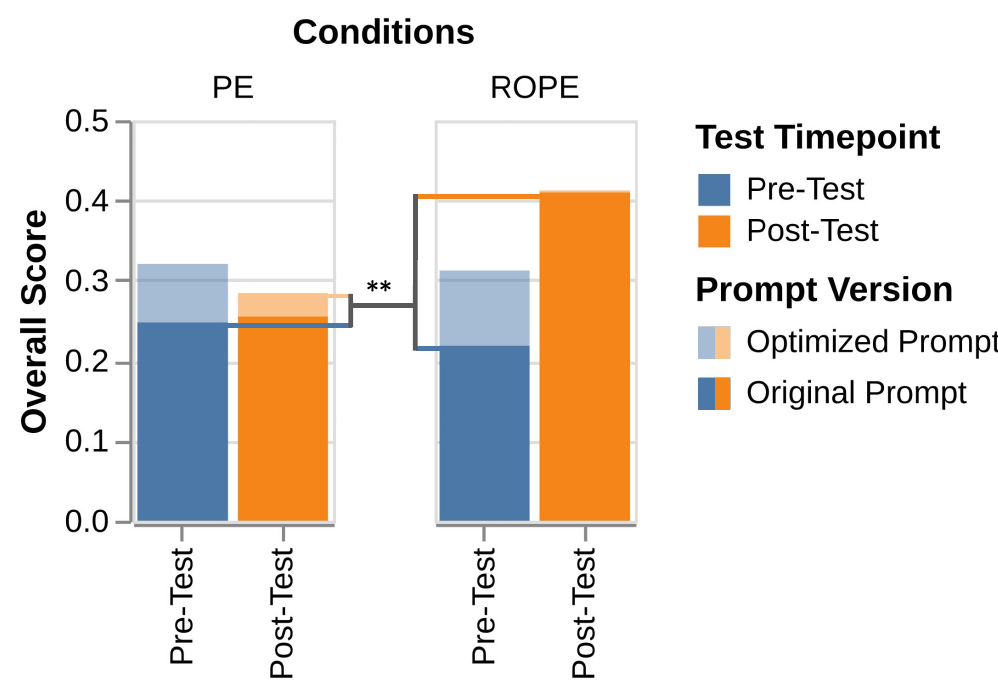}
    \caption{The pre-test and post-test overall scores between \ctrlgroup and \sysgroup conditions and the original and optimized prompt versions. \sysgroup's learning gain remains significantly higher than \ctrlgroup's optimized gains (** denotes $p \leq 0. 01$).}
    \label{fig:oriopt-gains}
    \vspace{-5pt}
    \Description{Bar chart comparing pre-test and post-test scores for original and optimized prompts in both PE and ROPE conditions. ROPE's original learning gains still outperform PE's optimized gains significantly (p = 0.01).}
\end{figure}

Importantly, \emph{the optimizer could not close the performance gap between the \sysgroup and \ctrlgroup group}.
We computed the gains of \ctrlgroup participants after using the optimizer ($7.3\%$ = optimized prompts' post-test scores $-$ original prompts' pre-test scores). 
Using a two-sample t-test, the \sysgroup's 19.1\% pre- to post-test gains remained significantly higher ($p=0.01$) than \ctrlgroup group's optimized gains ($7.3\%$), as shown in \cref{fig:oriopt-gains}.
This supports the finding that missing, highly customized requirements are unlikely to be added automatically.

While the optimizer generally had a positive impact, \emph{it might misinterpret user inputs or add irrelevant details}. For instance, for four out of five participants who requested an alphabetic list (A, B, C), the optimizer changed it to a numbered list. In some cases, these misinterpretations drastically reduced LLM performance. For example, two participants used the word ``interactive'' in their prompt, leading the optimizer to incorrectly add the role of ``intensive interactive web application developer'' and request the use of ``Flask and HTML'' in the optimized prompts, leading to poorly implemented code since it is much harder to generate a functional web app. 

These issues highlight potential \emph{gaps in user-expressed requirements and LLM-interpreted requirements}, suggesting that future training should address the risks of misinterpretation by either LLMs or optimizers (\cref{discuss:llm-rope}).
On a more positive note, such misinterpretations could provide useful feedback, alerting users to potential misalignment between their requirements and the LLM's understanding.

\paragraphBold{\sysname's effectiveness generalizes to more advanced (reasoning) LLMs.}
\label{subsec:res-o3mini}
Noticing the continuous evolvement of LLMs, we naturally ask the question: Would \sysname's effectiveness generalize beyond the \texttt{GPT-4o} we experimented, to models with better capabilities and alignments to human intents?
Recent advancements in LLMs have introduced reasoning models, which have improved instruction-following capabilities and exhibit better performance in complex reasoning tasks \cite{DeepSeek-AI2025-yd}. 
We conducted an additional experiment using OpenAI's \texttt{o3-mini}, a representative model of the emerging reasoning LLMs that has been shown to surpass \texttt{GPT-4o} \cite{Chiang2024-wj}. 
We had \texttt{o3-mini} generate outputs using the same prompts (both original and optimized ones as above), and repeated all the grading and analyses as done on \texttt{GPT-4o}.

The results showed that \emph{\sysname-trained prompts not only remained effective, but in many cases performed even better with \texttt{o3-mini}.}
Specifically, when using the overall scores calculated on \texttt{o3-mini} (which impacted the \emph{LLM Output Quality} score), we saw a larger learning gains compared to \texttt{GPT-4o} (23\% $>$ 19\% for the \sysgroup group, 2\% $>$ 1\% for \ctrlgroup group). 
This learning gap again could not be closed by the optimizer. 
Digging deeper, we found that \texttt{o3-mini} improved the \emph{LLM Output Quality} score by 9.7\% compared to \texttt{GPT-4o} (paired t-test, $p < 0.001$), which means that \texttt{o3-mini} better implemented the prompts.
This suggests that improved model capabilities do not render explicit human requirements obsolete, but rather amplify their benefits, making learning on requirements more reflective on the model output quality.

Furthermore, we also found a stronger alignment between user-specified requirements and generated outputs for \texttt{o3-mini}, indicated by the stronger correlation between \emph{Requirement Quality} and \emph{LLM Output Quality} scores when using \texttt{o3-mini} compared to \texttt{GPT-4o} ($\rho = 0.80 > 0.71$, as shown in \cref{tab:corr-by-task}). The largest improvement in correlation came from the \task{Connect4} task, where \texttt{o3-mini} showed better abilities to implement customized requirements like \req{Highlight the hovered column.} 
These findings underscore the continued relevance of human-authored requirements even as LLMs evolve (more discussions in \cref{discuss:llm-rope}). 

%% file: figures/table_correlations.tex
\begin{table}[]
\centering

\vspace{-5pt}
\caption{Spearman correlations ($\rho$) between LLM Output Quality and Requirement Quality scores by tasks and models.}
\vspace{-5pt}
\label{tab:corr-by-task}
\begin{tabular}{r r |r | r }
\toprule %
 &
 \textbf{Task} & \textbf{$\rho$ (GPT-4o)} & \textbf{$\rho$ (o3-mini)} \\ \midrule
\multirow{2}{*}{\rotatebox[origin=c]{90}{GPTs}} & 
           TripAdvisor     & 0.79       &  0.75 \\ 
          & OutlineAssistant& 0.75      &  0.83 \\ \midrule
\multirow{2}{*}{\rotatebox[origin=c]{90}{Game}} & 
           TicTacToe       & 0.90       &  0.90 \\ 
          & Connect4        & 0.10      &  0.67 \\ \midrule
\multicolumn{2}{r|}{\textbf{Overall}} & \textbf{0.71}   &  \textbf{0.80} \\ 
\bottomrule
\end{tabular}
\vspace{-10pt}
\end{table}

%% file: sections/discussion.tex
\section{Discussion}
\label{sec:discuss}

In this work, we introduce Requirement-Oriented Prompt Engineering (\sysname), a human-AI collaborative prompting paradigm where humans focus on effective requirements specification. Our evaluation shows that requirement-focused training significantly enhances novices' ability to extract and articulate requirements in prompts, leading to more goal-aligned LLM outputs; we also notice a key communication barrier between humans' under-specified requirements and LLMs' misinterpretations (\cref{subsec:res-optimizer}). 

While our work made an important step toward \sysname, several limitations remain, and there are many interesting questions to explore.
Here, we reflect on possible immediate extensions on requirement-focused training, as well as the longer-term evolution of the \sysname paradigm.

\subsection{Limitations and Next Steps on Requirement-Focused Training}
\label{subsec:limitation}

Our training program significantly improved participants' ability to extract and articulate requirements, though the skill remains challenging. Post-training assessments showed an average score of 41\% (max=63\%), and the top \sysgroup learner gained 43\% in scores from pre- to post-test. To further improve the effectiveness of requirement-focused training, we propose several next steps.

One easy extension is to make the training session longer. We observed that the 40-minute training time was difficult for non-native English speakers, and participants had limited opportunities for iteration due to time constraints. 
Future work should explore longer training sessions and adaptations for users with varying language proficiencies. 

In addition to extending length, our training can also be broadened for better generalizability.
Our study demonstrated that the requirement specification skills users acquired through training generalized to new GPTs and Game development tasks. 
However, an open question remains: how broadly do these requirement skills transfer to other natural language programming tasks (e.g., data analysis or creative content generation)? Does training on easier tasks generalize to harder tasks?
In preparation for generalizability, our \sysname interface is adaptable to different tasks, as the training and feedback loop can be easily generated based on provided reference requirements and LLM outputs. 
We open-source \sysname system and encourage its use for customized tasks across domains.
Moreover, beyond human-AI interactions, \sysname skills --- such as requirement extraction and iterative refinement --- can be broadly applicable to fields like software engineering and design. 
Future research should explore further transfer tasks in assessments, and also how \sysname skills may enhance communication, problem decomposition, and computational thinking across various fields.

Another dimension of generalizability is towards users' different prompting behaviors. 
Our study focused on extracting fixed requirements, which allowed for controlled feedback {and simulated the scenarios when interaction examples are available}. However, this setup did not fully capture all real-world prompting scenarios: 
users often engage in \emph{one-off} interactions (e.g., direct question-answering) or \emph{iterative and open-ended} prompting (e.g., chatting with LLMs on a rough idea)~\cite{Shankar2024-hy, Kim2024-lu}. 
Future studies should investigate more self-directed tasks, compare different types of prompting, and tailor training or support accordingly.
While providing feedback without clear requirements may be challenging, pre-generated materials on common novice approaches could offer useful scaffolding.

One primary bottleneck for broadening the scope is our lack of automatic assessment methods for the LLM-hard tasks. 
Our time-consuming manual grading effort limited our ability to conduct more experiments by e.g., grading multiple LLM outputs using different temperatures per user prompt, experimenting with multiple optimizers, etc. Future work should explore automatic assessments and their trade-offs.
For example, LLM-as-a judge paired with few-shot prompting might provide reasonable estimations~\cite{Zhao2022-tb} and might be suitable for tasks that contain substantial numbers of requirements where wrong grading on a single requirement does not drastically affect the assessment results. Future work can also explore alternative instructional designs such as using human experts or incorporate additional learning principles \cite{Koedinger2013-bp} to maximize the efficacy of \sysname training.

\subsection{Reflection on the \sysname Paradigm}

Along with prior work on requirement-oriented evaluation~\cite{Kim2024-lu}, we believe that \sysname moves us towards a future where \textbf{requirements serve as the central interface between AI and humans}
especially as models become more aligned with humans --- as we have seen in \cref{subsec:res-o3mini}, more advanced models better implement user requirements.
Here, we further discuss key skills humans and LLMs need to develop to prepare for a \sysname future. 
While our paper focused on a \sysname training system, our insights can also be useful for designing better prompting support tools. 

\paragraphBold{What should humans be equipped with for \sysname? End-to-end prompting skills}
\label{discuss:human-rope}

To ensure the success of \sysname, multiple skills need to be cultivated among humans.
Our training successfully helped users develop the ability to \emph{extract requirements from given examples and express them completely and correctly}, but our study also revealed more opportunities in requirement iteration and testing skills of an end-to-end prompting procedure. 
Specifically, users need to identify examples to compare LLM outputs and adjust the requirements accordingly. 

For example, users should \emph{create diverse and representative examples to identify and test requirements}, particularly for ambiguous or open-ended tasks involving edge cases or complex interactions. 
In our study, many users, especially those unfamiliar with games like \task{Connect4}, missed the testing scenarios like a tie game, which led them to omit the requirement \req{Display a "Tie Game!" message if no player wins} in their prompt.  
This is consistent to findings in debugging training \cite{Ma2024-ks} and other end-user prompt engineering work \cite{Zamfirescu-Pereira2023-is}, as novices often create too few test examples. 
Future studies should support users to generate more nuanced testing examples, supporting comprehensive hypothesis testing in prompt engineering \cite{Arawjo2024-pu}.

Additionally, users should \emph{iterate requirement granularity based on task ``LLM-hardness.''} This involves observing LLM failures and refining requirements, essentially developing a theory of mind for LLMs~\cite{10.1145/3613905.3636308}. 
Some of our participants were already thinking about \quoteinline{what is obvious enough for ChatGPT to already understand/perform well on} (S28) and adjust their prompt accordingly (although not necessarily correctly).
In our training, we introduced the concept of requirement specificity in \task{Tetris}, where users generate requirements in two levels of granularity (i.e., main steps and details, \cref{subsec:design-interaction}). 
However, some tasks required even greater specificity, such as generating non-standard \task{Connect4} gameplay elements \req{Cross out winning cells}. 
This ability to adjust specificity in requirements can be influenced by domain expertise, for example, expert developers can iterate prompts with more accurate details to successfully generate code while students often struggle \cite{Nam2024-td}. 
Future studies should offer adaptive support for adjusting requirement granularity based on task complexity and user expertise, such as via interactive dialog and automated diagnosis.

We believe that \emph{improving humans' requirement specification skills will remain important} despite rapid model advancements, since it is grounded in a fundamental challenge in human–LLM interaction: LLMs need access to user intent, but specific intent may not be available until users articulate it. 
As mentioned before, requirements might be ambiguous or entirely missing. 
The former might be mitigated by more capable models (or prompting support tools) that are better at e.g., making personalized guesses on user needs, asking clarification questions, or providing requirement-specific feedback~\cite{mu2023clarifygpt, zhang2023clarify}.
However, these can only be complementary to humans' own requirement elicitation --- a model that always asks for clarification on vague instructions can quickly become tedious and overwhelming to interact with.
More importantly, missing requirements will remain a challenge, especially in \emph{LLM-hard tasks}. When humans actually \emph{need to deviate the model's behavior but forget to articulate it}, models may already have a strong default behavior that prevents them from noticing anything missing. As a result, missing requirements risk becoming \emph{unknown unknowns}~\cite{lakkaraju2017identifying, rumsfeld2011known} to LLMs, and models may default to ``common sense'' completions that conflict with users' actual needs (e.g., defaulting to circles instead of X and O's in customized \task{Connect4} games).
Given decades of research in design and software engineering show that requirement elicitation is non-trivial for humans (\cref{subsec:RE}), we anticipate \sysname training will remain essential, even when LLMs and optimizers become more powerful.

\paragraphBold{What should optimizers and LLMs be capable of for \sysname? Support complementary task delegation}
\label{discuss:llm-rope}

For a successful \sysname future, optimizers and LLMs must develop complementary capabilities to support human skills.

First, as humans improve on articulating good requirements, optimizers should \textit{test different ``implementations'' of requirements}. This includes varied wordings, alternative expressions (e.g., zero-shot descriptions or few-shot examples), and the order, hierarchy, or strictness of requirement compositions. Optimizers should also dynamically group the same requirements presented in multiple formats to reveal which (combinations of) requirements are missing or ambiguous.

Second, optimizers can also \textit{assist test data synthesis}. More advanced optimizers need to iterate prompts on validation datasets and automatic objective functions, similar to machine learning. Insights from software engineering test automation or NLP approaches to extract or synthesize test cases from natural language requirements could be helpful \cite{Garousi2020-ne, yang-etal-2023-beyond, yu2024large, yang2024what}.

Finally, LLMs or optimizers should \emph{explicitly communicate their interpretations and reveal biases in their implementation}.
As we observed in our study, optimizers can misinterpret requirements or add unnecessary details (e.g., ``interactive game'' can lead to unwanted features like Flask integration). 
LLMs should explicitly communicate their understandings of user intent, such as translating ambiguous natural language requirements into more formal specifications \cite{Liu2023-hf, Lee2022-eo}, asking clarifying questions~\cite{Qian2024-ok, Lahiri2022-gb, Chen2024-du}, or extracting and refining generated criteria \cite{Kim2024-lu, Yuan2024-ce, Wang2024-xj}. 
This is not a perfect solution, as models' translations may still reflect their biases and models may miss to ask about \emph{LLM-hard} requirements. 
Nonetheless, to correctly implement requirements in LLM outputs when what is \emph{LLM-hard} keeps evolving, humans need to learn when and how to specify requirements with more granularity, and models need to keep communicating what requirements might be \emph{LLM-hard} or \emph{LLM-conflict}.